%% file: mfcs.tex
\documentclass[a4paper,USenglish,cleveref,autoref,thm-restate,authorcolumns,dvipsnames]{lipics-v2019}

\PassOptionsToPackage{prologue}{xcolor}

\newif\ifarxiv
\arxivtrue

\graphicspath{{./graphics/}}

\title{Algebra-based Loop Synthesis}
\titlerunning{Algebra-based Loop Synthesis}

\author{Andreas Humenberger}{TU Wien, Austria}{ahumenbe@forsyte.at}{}{}
\author{Laura Kov\'acs}{TU Wien, Austria \and Chalmers University of Technology, Sweden}{lkovacs@forsyte.at}{}{}
\authorrunning{A. Humenberger and L. Kov\'acs} 

\Copyright{Andreas Humenberger and Laura Kov\'acs} 

\ccsdesc[100]{Theory of computation~Program analysis}
\ccsdesc[100]{Theory of computation~Invariants}
\ccsdesc[100]{Computing methodologies~Symbolic and algebraic manipulation}

\keywords{loop synthesis, invariants, recurrence equations, non-linear arithmetic}

\ifarxiv
\else
\relatedversion{An extended version of the paper is available at \href{https://arxiv.org/abs/2004.11787}{\texttt{arxiv.org/abs/2004.11787}}.}
\fi

\supplement{}

\nolinenumbers 

\hideLIPIcs  

\EventEditors{John Q. Open and Joan R. Access}
\EventNoEds{2}
\EventLongTitle{42nd Conference on Very Important Topics (CVIT 2016)}
\EventShortTitle{CVIT 2016}
\EventAcronym{CVIT}
\EventYear{2016}
\EventDate{December 24--27, 2016}
\EventLocation{Little Whinging, United Kingdom}
\EventLogo{}
\SeriesVolume{42}
\ArticleNo{23}

\usepackage{booktabs}
\usepackage{subcaption}
\usepackage{amssymb}
\usepackage{amsmath}
\usepackage{xcolor}
\usepackage{multirow}
\usepackage{algorithm2e}
\usepackage{multicol}
\usepackage{mdframed}
\usepackage{listings}
\usepackage{tikz}
\usetikzlibrary{shapes.geometric, arrows}

\lstdefinelanguage{julia}
{
  keywordsprefix=\$,
  morekeywords={
    exit,whos,edit,load,is,isa,isequal,typeof,tuple,ntuple,uid,hash,finalizer,convert,promote,
    subtype,typemin,typemax,realmin,realmax,sizeof,eps,promote_type,method_exists,applicable,
    invoke,dlopen,dlsym,system,error,throw,assert,new,Inf,Nan,pi,im,begin,while,for,in,return,
    break,continue,macro,quote,let,if,elseif,else,try,catch,end,bitstype,ccall,do,using,module,
    import,export,importall,baremodule,immutable,local,global,const,Bool,Int,Int8,Int16,Int32,
    Int64,Uint,Uint8,Uint16,Uint32,Uint64,Float32,Float64,Complex64,Complex128,Any,Nothing,None,
    function,type,typealias,abstract
  },
  sensitive=true,
  morecomment=[l]{\#},
  morestring=[b]',
  morestring=[b]" 
}

\newcommand{\K}{\mathbb{K}}
\newcommand{\Q}{\mathbb{Q}}
\newcommand{\N}{\mathbb{N}}

\newcommand{\CC}{\mathcal{C}} 

\renewcommand{\vec}[1]{\boldsymbol{#1}}

\newcommand{\seq}[1]{\left({#1}\right)_{n=0}^\infty}

\newcommand{\kth}{\textsf{th}}
\newcommand{\ini}[1]{\bar{#1}}

\newcommand{\geom}{\omega}

\newcommand{\polysin}{\sqsubset}

\newcommand{\reserved}{\mathbf}
\newcommand{\WHILE}{\reserved{while}}
\newcommand{\DO}{\reserved{do}}
\newcommand{\END}{\reserved{end}}

\newcommand{\tool}[1]{\texttt{#1}}

\DeclareMathOperator{\cstr}{cstr}
\DeclareMathOperator{\decompose}{split}

\newcommand{\absynth}{\tool{Absynth}}
\newcommand{\absynthurl}{\url{https://github.com/ahumenberger/Absynth.jl}}

\theoremstyle{definition}
\newtheorem*{problem}{Problem}

\begin{document}

\maketitle

\begin{abstract}
  We present an algorithm for synthesizing program loops satisfying a given
  polynomial loop invariant. The class of loops we consider can be modeled by a
  system of algebraic recurrence equations with constant coefficients. We turn
  the task of loop synthesis into a polynomial constraint problem by precisely
  characterizing the set of all loops satisfying the given invariant. We prove
  soundness of our approach, as well as its completeness with respect to an a
  priori fixed upper bound on the number of program variables. Our work has
  applications towards program verification, as well as generating number
  sequences from algebraic relations. We implemented our work in the tool
  \absynth{} and report on our initial experiments with loop synthesis.
\end{abstract}

\section{Introduction}


The classical setting of
program synthesis has been to synthesize programs from proofs of logical
specifications that relate the inputs and the outputs 
of the program~\cite{MannaW80}. 
This traditional view of program synthesis has been refined to the setting of
syntax-guided synthesis (SyGuS)~\cite{Alur15}.  In addition to logical
specifications, SyGuS approaches consider further constraints on the program
template to be synthesized, thus limiting the search space of possible
solutions~\cite{GulwaniIJCAR16,GulwaniICSE10,DilligPLDI18,SolarICML19}. 

One of the main challenges in synthesis remains however to reason about program 
loops -- for example by answering the
question whether there exists a loop satisfying a given loop invariant and
synthesizing a loop with respect to a given invariant. We refer
to this task of synthesis as \emph{loop synthesis}, which can  
be considered as the reverse problem of loop invariant generation: rather
than 
generating invariants summarizing a given loop as in~\cite{Rodriguez-CarbonellK07,HumenbergerJK17,KincaidCBR18}, we synthesize loops whose
functional behavior is captured by a given invariant.

\begin{figure}[tb]
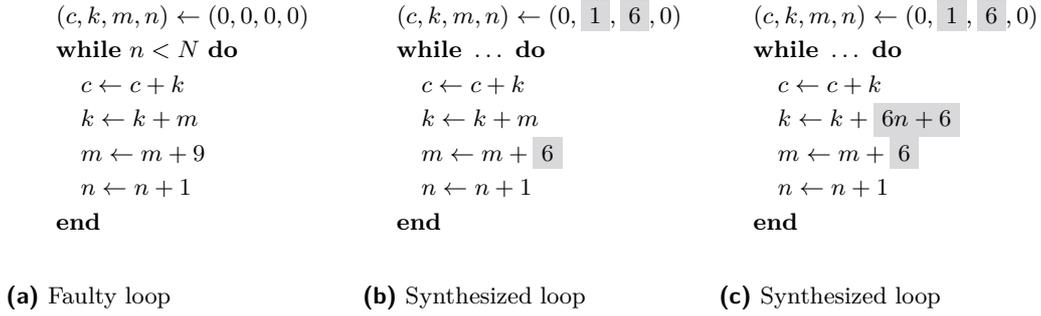

  \begin{subfigure}{.33\textwidth}
    \begin{center}
      \begin{tabular}{l}
      $(c, k, m, n) \gets (0, 0, 0, 0)$\\
      $\WHILE~n<N~\DO$\\
      \quad$c \gets c+k$\\
      \quad$k \gets k+m$\\
      \quad$m \gets m+9$\\
      \quad$n \gets n+1$\\
      $\END$
      \end{tabular}
    \end{center}
  \caption{Faulty loop}\label{fig:Dafny:a}
  \end{subfigure}\hfill
  \begin{subfigure}{.33\textwidth}
    \begin{center}
      \begin{tabular}{l}
      $(c, k, m, n) \gets (0, \colorbox{lipicsLightGray}{$1$}, \colorbox{lipicsLightGray}{$6$}, 0)$\\
      $\WHILE~\dots~\DO$\\
      \quad$c \gets c+k$\\
      \quad$k \gets k+m$\\
      \quad$m \gets m+\colorbox{lipicsLightGray}{$6$}$\\
      \quad$n \gets n+1$\\
      $\END$
      \end{tabular}
    \end{center}
  \caption{Synthesized loop}\label{fig:Dafny:b}
  \end{subfigure}\hfill
  \begin{subfigure}{.33\textwidth}
    \begin{center}
      \begin{tabular}{l}
      $(c, k, m, n) \gets (0, \colorbox{lipicsLightGray}{$1$}, \colorbox{lipicsLightGray}{$6$}, 0)$\\
      $\WHILE~\dots~\DO$\\
      \quad$c \gets c+k$\\
      \quad$k \gets k+\colorbox{lipicsLightGray}{$6n + 6$}$\\
      \quad$m \gets m+\colorbox{lipicsLightGray}{$6$}$\\
      \quad$n \gets n+1$\\
      $\END$
      \end{tabular}
    \end{center}
  \caption{Synthesized loop}\label{fig:Dafny:c}
  \end{subfigure}
  \caption{Program repair via loop
    synthesis. 
    Figures~\ref{fig:Dafny:b} and~\ref{fig:Dafny:c} are
    revised versions of Figure~\ref{fig:Dafny:a} such that
    ${c=n^3 \wedge k=3n^2+3n+1 \wedge m=6n+6}$ is an invariant of Figures~\ref{fig:Dafny:b}-\ref{fig:Dafny:c}.}
  \label{fig:Dafny}
\end{figure}

\subparagraph*{Motivating Example.} We motivate the use of loop
synthesis by 
considering the program snippet of Figure~\ref{fig:Dafny:a}. The loop in
Figure~\ref{fig:Dafny:a} is a variant of one of the examples from the online
tutorial\footnote{\url{https://rise4fun.com/Dafny/}} of the Dafny verification
framework~\cite{Dafny17}: the given program is not partially correct with
respect to the pre-condition $N\geq 0$ and post-condition $c=N^3$ and the task
is to revise/repair Figure~\ref{fig:Dafny:a} into a partially correct program
using the invariant $n \leq N \wedge c=n^3 \wedge k=3n^2+3n+1 \wedge m=6n+6$. 

Our work introduces an algorithmic approach to loop synthesis by relying on
algebraic recurrence equations and constraint solving over polynomials. In
particular, using our approach we automatically synthesize
Figures~\ref{fig:Dafny:b} and~\ref{fig:Dafny:c} by using the given non-linear
polynomial equalities $c=n^3 \wedge k=3n^2+3n+1 \wedge m=6n+6$ as input
invariant to our loop synthesis task. While we do not synthesize loop
guards, we note that we synthesize loops such that the given invariant holds for an
arbitrary (and thus unbounded) number of loop iterations. Both synthesized programs, with the
loop guard $n<N$ as in Figure~\ref{fig:Dafny:a},  revise Figure~\ref{fig:Dafny:a} into a partially correct
program with respect to the given requirements.

\subparagraph*{Algebra-based Loop Synthesis.} Following the SyGuS
setting, we consider 
additional requirements on the loop to be synthesized: we  impose
syntactic  requirements on the form of loop expressions and
guards. The imposed requirements allow us to 
\emph{reduce the synthesis task to
the problem of generating linear recurrences with constant
coefficients, called C-finite recurrences~\cite{KauersP11}}.
As such, we define \emph{our loop synthesis task} as follows:

\surroundwithmdframed[innertopmargin=1px, innerbottommargin=8px]{problem}
\begin{problem}[Loop Synthesis]
  Given a polynomial $p(\vec{x})$ over a set $\vec{x}$ of 
  variables, generate a loop $\mathcal{L}$ with program
  variables $\vec{x}$ such that
  \begin{enumerate}[(i)]
    \item ${p(\vec{x})=0}$ is an invariant of $\mathcal{L}$, and
    \item each program variable in $\mathcal{L}$ induces a C-finite number sequence.
  \end{enumerate}
\end{problem}



Our approach to synthesis is
conceptually different than other SyGuS-based methods, such
as~\cite{GulwaniIJCAR16,DilligPLDI18,SolarICML19}: rather than iteratively
refining both the input and the solution space of synthesized programs, we take
polynomial relations describing a potentially infinite set of input values and
precisely capture not just one loop, but the \emph{set of all loops} (i)~whose invariant is given by
our input polynomial and (ii)~whose variables induce C-finite number
sequences.
That is, any instance of this set yields a loop that is
partially correct by construction. Figures~\ref{fig:Dafny:b}
and~\ref{fig:Dafny:c} depict two solutions of our loop synthesis task for the
invariant ${c=n^3 \wedge k=3n^2+3n+1 \wedge m=6n+6}$.



The main steps of our approach are as follows. (i)~Let $p(\vec{x})$ be a polynomial
over variables $\vec{x}$ and let ${s\geq 0}$ be an upper bound on the number of
program variables occurring in the loop. If not specified, $s$ is considered to
be the number of variables from $\vec{x}$. 
%
%
%
(ii)~We use syntactic constraints over the loop body to be synthesized and
define a loop template, as given by our programming
model~\eqref{eq:loop}. Our programming model
imposes that 
 the functional behavior of the synthesized loops can
be modeled by a system of C-finite recurrences (Section~\ref{sec:model}). 
(iii)~By using the invariant property of $p(x)=0$ for the loops to the
synthesized, we construct a polynomial
constraint problem (PCP) characterizing the set of all
loops satisfying~\eqref{eq:loop} for which ${p(x) = 0}$
is a loop invariant (Section~\ref{sec:synth}). 
Our approach combines symbolic computation techniques over algebraic recurrence equations
with polynomial constraint solving. We prove that our approach to loop synthesis
is both \emph{sound} and \emph{complete}. By completeness we mean, that if
there is a loop $\mathcal{L}$ with at most $s$ variables satisfying the
invariant ${p(\vec{x})=0}$ such that the loop body meets our C-finite syntactic
requirements, then $\mathcal{L}$ is synthesized by our method (Theorem~\ref{thm:sound-complete}).
Moving beyond this a priori fixed bound $s$, that is, deriving an upper bound on
the number of program variables from the invariant, is an interesting but hard
mathematical challenge, with connections to the {inverse problem of
difference Galois theory}~\cite{Galois}.


We finally note that our work is
not restricted to specifications given by a single polynomial equality
invariant. Rather, the invariant given as input to our synthesis approach
can be conjunctions of polynomial equalities -- as also shown in
Figure~\ref{fig:Dafny}.

\subparagraph*{Beyond Loop Synthesis.}
Our work has potential applications beyond loop synthesis -- such as in
generating number sequences from algebraic relations and program optimizations.
\begin{itemize}
\item\label{usecase:seq}
\emph{Generating number sequences.} Our approach provides a partial solution to
an open mathematical problem: given a polynomial relation among number
sequences, e.g.
%
\begin{equation}
  \label{eq:fib:relation}
  f(n)^4 + 2f(n)^3f(n+1) - f(n)^2f(n+1)^2 - 2f(n)f(n+1)^3 + f(n+1)^4 = 1,
\end{equation}
synthesize algebraic recurrences defining these sequences. There exists no
complete method for solving this challenge, but we give a complete approach in
the C-finite setting parameterized by an a priori bound $s$ on the order of the
recurrences. For the above given relation among $f(n)$ and $f(n+1)$, our
approach generates the C-finite recurrence equation $f(n+2)=f(n+1)+f(n)$ which
induces the Fibonacci sequence.

\item\label{usecase:opt}
\emph{Program optimizations.} Given a polynomial invariant, our approach
generates a PCP such that any solution to this PCP yields a loop
satisfying the given invariant. 
By using additional constraints
encoding a cost function on the loops to be synthesized, our method can be
extended to synthesize loops that are optimal with respect to the considered
costs, for example  synthesizing loops that use only addition in
variable updates. 
Consider for example Figures~\ref{fig:Dafny:b}-\ref{fig:Dafny:c}: 
the loop body of Figure~\ref{fig:Dafny:b} uses only addition, whereas
Figure~\ref{fig:Dafny:c} implements also multiplications by
constants. 

\end{itemize}

\subparagraph*{Contributions.}
In summary, this paper makes the following contributions.
\begin{itemize}
\item
We propose an \emph{automated procedure for synthesizing loops} that are
partially correct with respect to a given polynomial loop invariant
(Section~\ref{sec:synth}). 
By exploiting properties of C-finite sequences, we construct a PCP which
precisely captures \emph{all solutions} of our loop synthesis task. We are not
aware of other approaches synthesizing loops from (non-linear) polynomial
invariants. 



\item
We prove that our approach to loop synthesis is sound and complete
(Theorem~\ref{thm:sound-complete}). That is, if there is a loop whose invariant
is captured by our given specification, our approach synthesizes this loop. To
this end, we consider completeness modulo an a priori fixed upper bound $s$ on
the number of loop variables.

%

%
\item 
We implemented our approach in the new open-source framework \absynth{}. We evaluated our work on a number of
academic examples and considered measures for handling the solution space of
loops to be synthesized (Section~\ref{sec:synth:opt}).

\end{itemize}


\section{Preliminaries}
\label{sec:prelim}

Let $\K$ be a computable field with characteristic zero. We also
assume $\K$ to be 
algebraically closed, that is, every non-constant polynomial in $\K[x]$ has at
least one root in $\K$. 
The algebraic closure $\bar{\Q}$ of the field of rational numbers $\Q$ 
is such a field; $\bar{\Q}$ is called the field of algebraic numbers. 

Let $\K[x_1,\dots,x_n]$ denote the multivariate polynomial ring with
variables $x_1,\dots,x_n$.  For a list $x_1,\dots,x_n$, we write $\vec{x}$ if the
number of variables is known from the context or irrelevant. As $\K$ is
algebraically closed, every polynomial $p\in\K[\vec{x}]$ of degree
$r$ has exactly $r$ roots.
\ifarxiv 
Therefore, the following theorem follows immediately:
\begin{theorem}
  \label{thm:inifite-zeros}
  The zero polynomial is the only polynomial in $\K[\vec{x}]$ having infinitely
  many roots.
\end{theorem}
\fi 




\subsection{Polynomial Constraint Problem (PCP)}

A \emph{polynomial constraint} $F$ is a constraint of the form $p \bowtie 0$
where $p$ is a polynomial in $\K[\vec{x}]$ and $\mathord{\bowtie} \in
\{<,\leq,=,\neq,\geq,>\}$. A \emph{clause} is then a disjunction ${C = F_1 \lor
\dots \lor F_m}$ of polynomial constraints. A \emph{unit clause} is a special
clause consisting of a single disjunct (i.e. $m=1$). A \emph{polynomial
constraint problem (PCP)} is then given by a set of clauses $\CC$. We say that a
variable assignment $\sigma : \{x_1,\dots,x_n\} \rightarrow \K$ satisfies a
polynomial constraint $p\bowtie0$ if $p(\sigma(x_1),\dots,\sigma(x_n)) \bowtie
0$ holds. Furthermore, $\sigma$ satisfies a clause $F_1 \lor \cdots \lor F_m$ if
for some $i$, $F_i$ is satisfied by $\sigma$. Finally, $\sigma$ satisfies a
clause set -- and is therefore a solution of the PCP -- if every clause within
the set is satisfied by $\sigma$. We write $\CC\polysin\K[\vec{x}]$ to indicate
that all polynomials in the clause set $\CC$ are contained in $\K[\vec{x}]$. For
a matrix $M$ with entries $m_1,\dots,m_s$ we define the clause set $\cstr(M)$ to
be ${\{m_1=0,\dots,m_s=0\}}$.


\subsection{Number Sequences and Recurrence Relations}
\label{sec:cfinite}

A sequence $\seq{x(n)}$ is called \emph{C-finite} if it satisfies a linear
recurrence with constant coefficients, also known as C-finite recurrence~\cite{KauersP11}. Let
$c_0,\dots,c_{r-1} \in\K$ and $c_0\neq0$, then
\begin{equation}
  \label{eq:cfinite-form}
  x(n+r) + c_{r-1}x(n+r-1) + \cdots + c_1x(n+1) + c_0x(n) = 0
\end{equation}
is a C-finite recurrence of \emph{order} $r$. The order of a sequence is defined
by the order of the recurrence it satisfies. We refer to a recurrence
of order $r$ also as an $r$-order recurrence, for example as a first-order
recurrence when $r=1$ or a second-order recurrence when $r=2$. 
A recurrence of order $r$ and $r$ initial values define a sequence, and
different initial values lead to different sequences. For simplicity, we write
$\seq{x(n)} = 0$ for $\seq{x(n)} = \seq{0}$.

\begin{example}
  Let $a\in\K$. The constant sequence $\seq{a}$ satisfies a first-order
  recurrence equation ${x(n+1) = x(n)}$ with ${x(0)=a}$. The geometric sequence
  $\seq{a^n}$ satisfies ${x(n+1) = a x(n)}$ with ${x(0)=1}$. The sequence
  $\seq{n}$ satisfies a second-order recurrence ${x(n+2) = 2 x(n+1) - x(n)}$
  with ${x(0)=0}$ and ${x(1)=1}$.\qed
\end{example}

From the closure properties of C-finite sequences~\cite{KauersP11}, the product and
the sum of C-finite sequences are also
C-finite. Moreover, we also have the following properties:

\ifarxiv 

\begin{theorem}[\cite{KauersP11}]
  \label{thm:cfinite-closure}
  Let $\seq{u(n)}$ and $\seq{v(n)}$ be C-finite sequences of order $r$ and $s$,
  respectively. Then:
  \begin{enumerate}
    \item $\seq{u(n) + v(n)}$ is C-finite of order at most $r+s$, and
    \item $\seq{u(n)\cdot v(n)}$ is C-finite of order at most $rs$.\qed
  \end{enumerate}
\end{theorem}

\begin{theorem}[\cite{KauersP11}]
  \label{thm:cfinite-zero}
  Let $\geom_1,\dots,\geom_t\in\K$ be pairwise distinct and
  $p_1,\dots,p_t\in\K[x]$. The sequence $\seq{p_1(n)\geom_1^n + \cdots +
  p_t(n)\geom_t^n}$ is the zero sequence if and only if the sequences
  $\seq{p_1(n)},\dots,\seq{p_t(n)}$ are zero.\qed
\end{theorem}

\fi 

\begin{theorem}[\cite{KauersP11}]
  \label{thm:poly-zero}
  Let $p = c_0 + c_1 x + \cdots + c_kx^k \in\K[x]$. Then
  ${\seq{p(n)} = 0}$ if and only if $c_0 = \cdots = c_k = 0$.\qed
\end{theorem}

\begin{theorem}[\cite{KauersP11}]
  \label{thm:finite-init-values}
  Let $\seq{u}$ be a sequence satisfying a C-finite recurrence of order $r$.
  Then, $u(n) = 0$ for all $n\in\N$ if and only if $u(n) = 0$ for
  $n\in \{0,\dots,r-1\}$.\qed
\end{theorem}

We define a \emph{system of C-finite recurrences} of \emph{order} $r$ and \emph{size} $s$
to be of the form
\begin{equation*}
  X_{n+r} + C_{r-1}X_{n+r-1} + \cdots + C_1X_{n+1} + C_0X_n = 0
\end{equation*}
where $X_n = \begin{pmatrix} x_1(n) & \cdots & x_s(n) \end{pmatrix}^\intercal$
and $C_i\in\K^{s\times s}$. Every C-finite recurrence system can be transformed
into a first-order system of recurrences by increasing the size such
that we get
\begin{equation}
  \label{eq:cfinite-rec}
  X_{n+1} = B X_{n} \qquad \text{where $B$ is invertible.}
\end{equation}
The closed form solution of a C-finite recurrence system~(\ref{eq:cfinite-rec}) is determined by the roots $\geom_1,\dots, \geom_t$ of the
characteristic polynomial of $B$, or equivalently by the eigenvalues
$\geom_1,\dots, \geom_t$ of $B$.
We recall that the
characteristic polynomial $\chi_B$ of the matrix $B$ is defined as
$\chi_B(\geom) = \det(\geom I - B)$, where $\det$ denotes the (matrix)
determinant and $I$ the identity matrix.
Let $m_1,\dots,m_t$ respectively denote the 
multiplicities of the roots $\geom_1,\dots, \geom_t$ of $\chi_B$. 
%
The closed form of~\eqref{eq:cfinite-rec} is then given by
\begin{equation}
  \label{eq:cfinite-cf}
  X_n = \sum_{i=1}^t \sum_{j=1}^{m_i} C_{ij} \geom_i^n n^{j-1} \qquad \text{with $C_{ij} \in \K^{s\times 1}$.}
\end{equation}
However, not every choice of the $C_{ij}$
gives rise to a solution. For obtaining a solution, we substitute the general
form~(\ref{eq:cfinite-cf}) into the original system~(\ref{eq:cfinite-rec}) and
compare coefficients. \ifarxiv The following example illustrates the procedure for
computing closed form solutions.\fi

\ifarxiv 

\begin{example}
  \label{ex:fibonacci}
  The most well-known C-finite sequence is the Fibonacci sequence satisfying a
  recurrence of order $2$ which corresponds to the following first-order
  recurrence system:
  \begin{equation}
    \label{eq:ex:cfinite1}
    \begin{pmatrix}
      f(n+1) \\
      g(n+1)
    \end{pmatrix}=
    \begin{pmatrix}
      1 & 1 \\
      1 & 0
    \end{pmatrix}
    \begin{pmatrix}
      f(n) \\
      g(n)
    \end{pmatrix}
  \end{equation}
  The eigenvalues of $B$ are given by $\geom_{1,2}=\frac{1}{2}(1 \pm
  \sqrt{5})$ with multiplicities $m_1 = m_2 = 1$. Therefore, the general
  solution for the recurrence system is of the form
  \begin{equation}
    \label{eq:ex:cfinite2}
    \begin{pmatrix}
      f(n) \\
      g(n)
    \end{pmatrix}=
    \begin{pmatrix}
      c_1 \\
      c_2
    \end{pmatrix}\geom_1^n +
    \begin{pmatrix}
      d_1 \\
      d_2
    \end{pmatrix}\geom_2^n
    .
  \end{equation}
  By substituting \eqref{eq:ex:cfinite2} into \eqref{eq:ex:cfinite1}, we get the
  following constraints over the coefficients:
  \begin{equation*}
    \begin{pmatrix}
      c_1 \\
      c_2
    \end{pmatrix}\geom_1^{n+1} +
    \begin{pmatrix}
      d_1 \\
      d_2
    \end{pmatrix}\geom_2^{n+1}
    =
    \begin{pmatrix}
      1 & 1 \\
      1 & 0
    \end{pmatrix}
    \left(
    \begin{pmatrix}
      c_1 \\
      c_2
    \end{pmatrix}\geom_1^n +
    \begin{pmatrix}
      d_1 \\
      d_2
    \end{pmatrix}\geom_2^n
    \right)
  \end{equation*}
  Bringing everything to one side yields:
  \begin{equation*}
    \begin{pmatrix}
      c_1\geom_1 - c_1 - c_2 \\
      c_2\geom_1 - c_1
    \end{pmatrix}\geom_1^{n} +
    \begin{pmatrix}
      d_1\geom_2 - d_1 - d_2\\
      d_2\geom_2 - d_1
    \end{pmatrix}\geom_2^{n}
    = 0
  \end{equation*}
  For the above equation to hold, the coefficients of the $\geom_i^n$ have to be
  $0$. That is, the following linear system determines $c_1,c_2$ and $d_1,d_2$:
  \begin{equation*}
    \begin{pmatrix}
      \geom_1 - 1 & -1 & 0 & 0 \\
      -1 & \geom_1 & 0 & 0 \\
      0 & 0 & \geom_2 - 1 & -1 \\
      0 & 0 & -1 & \geom_2
    \end{pmatrix}
    \begin{pmatrix}
      c_1 \\
      c_2 \\
      d_1 \\
      d_2
    \end{pmatrix}
    = 0
  \end{equation*}
  The solution space is generated by $(\geom_1,1,0,0)$ and $(0,0,\geom_2,1)$.
  The solution space of the C-finite recurrence system hence consists of linear
  combinations of
  \begin{equation*}
    \begin{pmatrix}
      \geom_1 \\
      1
    \end{pmatrix}\geom_1^n
    \quad\text{and}\quad
    \begin{pmatrix}
      \geom_2 \\
      1
    \end{pmatrix}\geom_2^n.
  \end{equation*}
  That is, by solving the linear system
  \begin{align*}
    \begin{pmatrix}
      f(0) \\
      g(0)
    \end{pmatrix}&=
    E
    \begin{pmatrix}
      \geom_1 \\
      1
    \end{pmatrix}\geom_1^0 +
    F
    \begin{pmatrix}
      \geom_2 \\
      1
    \end{pmatrix}\geom_2^0\\
    \begin{pmatrix}
      f(1) \\
      g(1)
    \end{pmatrix}=
    \begin{pmatrix}
      1 & 1 \\
      1 & 0
    \end{pmatrix}
    \begin{pmatrix}
      f(0) \\
      g(0)
    \end{pmatrix}&=
    E
    \begin{pmatrix}
      \geom_1 \\
      1
    \end{pmatrix}\geom_1^1 +
    F
    \begin{pmatrix}
      \geom_2 \\
      1
    \end{pmatrix}\geom_2^1
  \end{align*}
  for ${E, F\in\K^{2\times 1}}$ with $f(0)=1$ and $g(0)=0$, we get
  closed forms for~\eqref{eq:ex:cfinite1}: 
  \begin{equation*}
    f(n) = \frac{5+\sqrt{5}}{5(1+\sqrt{5})}\geom_1^{n+1} - \frac{1}{\sqrt{5}}\geom_2^{n+1}
    ~\text{and}~
    g(n) = \frac{1}{\sqrt{5}}\geom_1^{n} - \frac{1}{\sqrt{5}}\geom_2^{n}
  \end{equation*}
  Then $f(n)$ represents the Fibonacci sequence starting at $1$ and $g(n)$
  starts at $0$. Solving for $E$ and $F$ with symbolic $f(0)$ and $g(0)$ yields
  a parameterized closed form, where the entries of $E$ and $F$ are linear
  functions in the symbolic initial values.
\end{example}

\fi 


\section{Our Programming Model}\label{sec:model}

Given a polynomial relation ${p(x_1,\dots,x_s)=0}$, our loop synthesis procedure
generates a first-order C-finite recurrence system of the
form~\eqref{eq:cfinite-rec} with ${X_n = \begin{pmatrix} x_1(n) & \cdots &
x_s(n) \end{pmatrix}^\intercal}$, such that ${p(x_1(n),\dots,x_s(n))=0}$ holds
for all ${n\in\N}$. It is not hard to argue that every first-order C-finite
recurrence system corresponds to a loop with simultaneous variable assignments
of the following form:
%
%
\begin{equation}
  \label{eq:loop}
  \begin{tabular}{l}
  $(x_1,\dots,x_s) \gets (a_1,\dots,a_s)$\\
  $\WHILE~true~\DO$\\
  \quad$(x_1,\dots,x_s) \gets (p_1(x_1,\dots,x_s),\dots,p_s(x_1,\dots,x_s))$\\
  $\END$
  \end{tabular}
\end{equation}
The program variables $x_1,\dots,x_s$ are numeric, $a_1,\dots,a_s$ are
(symbolic) constants in $\K$ and $p_1,\dots,p_s\in\K[x_1,\dots,x_s]$. For every
loop variable $x_i$, we denote by $x_i(n)$ the value of $x_i$ at the $n$th loop
iteration. That is, we view loop variables $x_i$ as sequences $\seq{x_i(n)}$. 


We call a loop~\eqref{eq:loop} \emph{parameterized} if at
least one of $a_1,\dots,a_s$ is symbolic, and \emph{non-parameterized}
otherwise.



\ifarxiv 
\begin{remark}
  While the output of our synthesis procedure is basically an affine program, we
  note that C-finite recurrence systems capture a larger class of programs.
  E.g.~the program:
  \[(x,y) \gets (0,0);~\WHILE~true~\DO~(x,y) \gets (x+y^2,y+1)~\END\]
  can be modeled by a C-finite recurrence system of order $4$, which can be
  turned into an equivalent first-order system of size $6$. That is, in order to
  synthesize a program which induces the sequences $\seq{x(n)}$ and $\seq{y(n)}$
  we have to consider a recurrence system of size $6$.\qed
\end{remark}
\fi 


\ifarxiv 
\begin{example}
  \label{ex:fibonacci-loop}
  The recurrence system~\eqref{eq:ex:cfinite1} in Example~\ref{ex:fibonacci}
  corresponds to the following loop:
  \[(f,g) \gets (1,0);~\WHILE~true~\DO~(f,g) \gets (f+g,f)~\END \hfill\qed\]
\end{example}
\fi 


\subparagraph*{Algebraic relations and loop invariants.} 
Let $p$ be a polynomial in $\K[z_1,\dots,z_s]$ and let
$(x_1(n))^\infty_{n=0},\dots,(x_s(n))^\infty_{n=0}$ be number sequences. We call
$p$ an \emph{algebraic relation} for the given sequences if
${p(x_1(n),\dots,x_s(n)) = 0}$ for all ${n\in\N}$. Moreover, $p$ is an algebraic
relation for a system of recurrences if it is an algebraic relation for the
corresponding sequences. It is immediate that for every algebraic relation $p$
of a recurrence system, ${p=0}$ is a \emph{loop invariant} for the corresponding
loop~\eqref{eq:loop}; that is, ${p=0}$ holds before and after every
loop iteration.


\section{Algebra-based Loop Synthesis}
\label{sec:synth}

We now present our approach for synthesizing loops satisfying a given polynomial
property (invariant). We transform the loop synthesis problem into
a PCP as described in 
Section~\ref{sec:synth:overview}. In Section~\ref{sec:synth:nonparam}, we
introduce the clause sets of our PCP which precisely describe the solutions for the
synthesis of loops, in particular to  non-parameterized loops. 
\ifarxiv
We extend this approach in Section~\ref{sec:synth:param} to parameterized loops.
\fi


\newcommand{\Croots}{\CC_\mathsf{roots}}
\newcommand{\Ccoeff}{\CC_\mathsf{coeff}}
\newcommand{\Cinit}{\CC_\mathsf{init}}
\newcommand{\Calg}{\CC_\mathsf{alg}}

\subsection{Setting and Overview of Our Method}
\label{sec:synth:overview}

Given a constraint ${p=0}$ with $p\in\K[x_1,\dots,x_s,y_1,\dots,y_s]$,
we aim 
to synthesize a system of C-finite recurrences such that $p$ is an algebraic
relation thereof. Intuitively, the values of loop variables $x_1,\dots,x_s$ are
described by  the number sequences $x_1(n),\dots,x_s(n)$ for arbitrary $n$, and
$y_1,\dots,y_s$ correspond to the initial values $x_1(0),\dots,x_s(0)$. That is,
we have a polynomial relation $p$ among loop variables $x_i$
and their initial values $y_i$, for which we synthesize a loop~\eqref{eq:loop} such that ${p=0}$
is a loop invariant of loop~\eqref{eq:loop}.

\begin{remark}
  Our approach is not limited to invariants describing the relationship between
  program variables among a single loop iteration. Instead, it naturally extends
  to relations among different loop iterations. For instance, by considering the
  relation in equation~\eqref{eq:fib:relation}, we synthesize a loop computing
  the Fibonacci sequence.
\end{remark}

\begin{figure}
  \centering
  \tikzstyle{inner} = [rectangle, rounded corners=1mm, inner xsep=3mm, text width=2cm, dashed, minimum width=2.0cm, minimum height=1cm, text centered, draw=black] 
  \tikzstyle{input} = [rectangle, rounded corners=1mm, inner xsep=3mm, text width=2cm, minimum width=2.0cm, minimum height=1cm, text centered, draw=black] 
  \tikzstyle{arrow} = [thick,->,>=stealth]
  \begin{tikzpicture}[node distance=2cm]
    \node (cfsys) [inner] {Closed form system};
    \node (poly) [input, left of=cfsys, xshift=-1.8cm] {Polynomial invariant};
    \node (recsys) [inner, right of=cfsys, xshift=2.9cm] {Recurrence system};
    \node (loop) [right of=recsys, xshift=1.4cm] {Loop};
  
    \draw (poly) -- node[anchor=south] {$\Calg$} (cfsys);
    \draw [arrow] (recsys) -- (loop);
    \draw (cfsys) -- node[anchor=south] {$\Croots$, $\Ccoeff$} node[anchor=north] {$\Cinit$} (recsys);
  \end{tikzpicture}
  \caption{Overview of the PCP 
    describing loop synthesis}
  \label{fig:overview}
\end{figure}
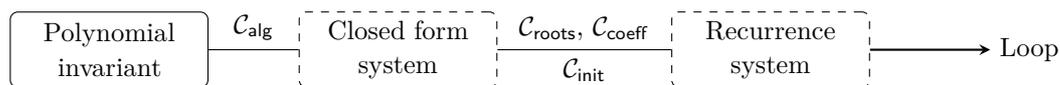

The key step in our work comes with precisely capturing the solution
space for our loop synthesis problem as a PCP.
Our PCP is divided into the clause sets $\Croots$, $\Ccoeff$, $\Cinit$
and $\Calg$, as illustrated in Figure~\ref{fig:overview} 
and explained next. Our PCP implicitly describes a first-order C-finite
recurrence system and its corresponding closed form system. The one-to-one
correspondence between these two systems is captured by the clause sets
$\Croots$, $\Ccoeff$ and $\Cinit$. Intuitively, these constraints mimic the
procedure for computing the closed form of a recurrence system
(see~\cite{KauersP11}). The clause set $\Calg$ interacts between the closed form
system and the polynomial constraint ${p=0}$, and ensures that $p$ is an
algebraic relation of the system. Furthermore, the recurrence system is
represented by the matrix $B$ and the vector $A$ of initial values where both
consist of symbolic entries. Then a solution of our PCP -- which assigns values
to those symbolic entries -- yields a desired synthesized loop.

In what follows we only consider a unit constraint $p=0$ as input to our loop
synthesis procedure. However, our approach naturally extends to conjunctions of polynomial
equality constraints.




\subsection{Synthesizing Non-Parameterized Loops}
\label{sec:synth:nonparam}

We now present our work for synthesizing loops, in particular 
non-parameterized loops~\eqref{eq:loop}. That is, we aim at computing concrete initial values
for all program variables. Our implicit representation of the recurrence system
is thus of the form 
\begin{equation}
  \label{eq:nonparam:recsys}
  X_{n+1} = BX_n \qquad X_0 = A
\end{equation}
where $B\in\K^{s\times s}$ is invertible and $A\in\K^{s\times 1}$, both
containing symbolic entries.

As described in Section~\ref{sec:cfinite}, the closed form
of~\eqref{eq:nonparam:recsys}
is determined by the eigenvalues $\geom_i$ of $B$ which we thus need to
synthesize. 
Note that $B$ may contain both symbolic and concrete
values. Let us denote the symbolic entries of $B$ by $\vec{b}$. 
Since $\K$ is algebraically
closed we know that $B$ has $s$ (not necessarily distinct) eigenvalues.
We therefore fix a set of distinct
symbolic eigenvalues $\geom_1,\dots,\geom_t$ together with their multiplicities
$m_1,\dots,m_t$ with $m_i>0$ for $i=1,\dots,t$ such that $\sum_{i=1}^t m_i = s$. 
We call $m_1,\dots,m_t$ an \emph{integer partition} of $s$.
We next define the clause sets of our PCP. 

\subparagraph*{Root constraints $\Croots$.}

The clause set $\Croots$ imposes that $B$ is invertible and ensures
that $\geom_1,\dots,\geom_t$ are distinct symbolic eigenvalues with
multiplicities $m_1,\dots,m_t$. 
Note that $B$ is invertible if and only if all eigenvalues $\geom_i$ are non-zero.
Furthermore, since $\K$ is algebraically closed, every polynomial $f(z)$ can be
written as the product of linear factors of the form $z - \geom$, with
$\geom\in\K$, such that $f(\geom)=0$. Therefore, the equation 
\begin{equation*}
  \chi_B(z) = (z - \geom_1)^{m_1} \cdots (z - \geom_t)^{m_t}
\end{equation*}
holds for all $z\in\K$, where $\chi_B(z)\in\K[\vec{\geom},\vec{b},z]$.
Bringing everything to one side, we get 
\begin{equation*}
  q_0 + q_1z + \cdots + q_d z^d = 0, 
\end{equation*}
implying that the ${q_i\in\K[\vec{\geom},\vec{b}]}$ have to be zero. 
The clause set characterizing the eigenvalues $\geom_i$ of $B$ is then
\begin{align*}
  \Croots = \{ q_0=0,\dots,q_d=0 \}
  \cup \bigcup_{\substack{i,j=1,\dots,t\\i\neq j}} \{\geom_i \neq
  \geom_j\} \cup \bigcup_{i=1,\dots,t} \{\geom_i \neq 0\}.
\end{align*}
%

\subparagraph*{Coefficient constraints $\Ccoeff$.}

The fixed symbolic roots/eigenvalues $\geom_1,\dots,\geom_t$ with multiplicities
$m_1,\dots,m_t$ induce the general closed form solution
\begin{equation}
  \label{eq:generalform}
  X_n = \sum_{i=1}^t \sum_{j=1}^{m_i} C_{ij} \geom_i^n n^{j-1}
\end{equation}
where the ${C_{ij}\in\K^{s\times 1}}$ are column vectors containing symbolic
entries. As stated in Section~\ref{sec:cfinite}, not every choice of
the $C_{ij}$ gives rise to a valid solution. Instead,  $C_{ij}$ have to obey certain
conditions which are determined by substituting into the original recurrence
system of \eqref{eq:nonparam:recsys}:
\begin{align*}
  X_{n+1} 
  &= \sum_{i=1}^t \sum_{j=1}^{m_i} C_{ij} \geom_i^{n+1}(n+1)^{j-1}
  = \sum_{i=1}^t \sum_{j=1}^{m_i} \left(\sum_{k=j}^{m_i} \binom{k-1}{j-1} C_{ik} \geom_i\right) \geom_i^{n}n^{j-1} \\
  &= B \left(\sum_{i=1}^t \sum_{j=1}^{m_i} C_{ij} \geom_i^n n^{j-1}\right) = BX_n
\end{align*}
Bringing everything to one side yields ${X_{n+1} - BX_n = 0}$ and thus
\begin{equation}
  \label{eq:cstr-cf}
  \sum_{i=1}^t \sum_{j=1}^{m_i} \underbrace{\left(\left(\sum_{k=j}^{m_i} \binom{k-1}{j-1} C_{ik} \geom_i\right) - BC_{ij}\right)}_{D_{ij}} \geom_i^{n}n^{j-1} = 0.
\end{equation}
Equation~\eqref{eq:cstr-cf} holds for all $n\in\N$. By 
Theorem~\ref{thm:poly-zero} we then have $D_{ij} = 0$ for all $i,j$
and define  
%
%
\begin{equation*}
  \Ccoeff = \bigcup_{i=1}^t \bigcup_{j=1}^{m_i} \cstr(D_{ij}).
\end{equation*}
%

\subparagraph*{Initial values constraints $\Cinit$.}
The constraints $\Cinit$ describe properties of initial values
$x_1(0),\dots,x_s(0)$. We enforce that \eqref{eq:generalform} equals
$B^nX_0$, for $n=0,\dots,d-1$, where $d$ is the degree of the characteristic
polynomial $\chi_B$ of $B$, by 
\begin{equation*}
  \label{eq:initvalues}
  \Cinit = \cstr(M_0) \cup\cdots\cup \cstr(M_{d-1})
\end{equation*}
where $M_i = X_i - B^iX_0$, with $X_0$ as in~\eqref{eq:nonparam:recsys} and $X_i$
being the right-hand side of~\eqref{eq:generalform} where $n$ is replaced by
$i$.
%

\subparagraph*{Algebraic relation constraints $\Calg$.}
The constraints $\Calg$ are defined to ensure that $p$ is an
algebraic relation among the $x_i(n)$. Using~\eqref{eq:generalform}, the
closed forms of the $x_i(n)$ are expressed as
\begin{equation*}
  \label{eq:cfpoly}
  x_i(n) = p_{i,1} \geom_1^n + \cdots + p_{i,t} \geom_t^n
\end{equation*}
where the $p_{i,j}$ are polynomials in $\K[n,\vec{c}]$. By substituting the
closed forms and the initial values into the polynomial $p$, we get
\begin{equation}
  \label{eq:invpoly}
  \begin{aligned}
  p' 
  = p(x_1(n),\dots,x_s(n),x_1(0),\dots,x_s(0)) 
  = q_0 + nq_1 + n^2q_2 + \cdots + n^kq_k
  \end{aligned}
\end{equation}
where the $q_i$ are of the form
\begin{equation}
  \label{eq:qform}
  w_{i,1}^nu_{i,1} + \cdots + w_{i,{\ell}}^n u_{i,{\ell}}
\end{equation}
with $u_{i,1},\dots,u_{i,{\ell}}\in\K[\vec{a},\vec{c}]$ and
$w_{i,1},\dots,w_{i,{\ell}}$ being monomials in $\K[\vec{\geom}]$. 

\begin{proposition}
  \label{prop:coeff-zero}
  Let $p$ be of the form~\eqref{eq:invpoly}. Then ${\seq{p(n)}=0}$ if and
  only if ${\seq{q_i(n)} = 0}$ for $i=0,\dots,k$. \qed
\end{proposition}

\ifarxiv 
\begin{proof}
  One direction is obvious and for the other assume ${p(n)=0}$. By rearranging
  $p$ we get $p_1(n) w_1^n + \cdots + p_\ell(n) w_\ell^n$. Let
  $\tilde{\geom}_1,\dots,\tilde{\geom}_t\in\K$ be such that $\tilde{p} = p_1(n)
  \tilde{w}_1^n + \cdots + p_\ell(n) \tilde{w}_\ell^n = 0$ with $\tilde{w}_i =
  w_i(\vec{\tilde\geom})$. Note that the $\tilde{w}_i$ are not necessarily
  distinct. However, consider $v_1,\dots,v_r$ to be the pairwise distinct elements
  of the $\tilde{w}_i$. Then we can write $\tilde{p}$ as $\sum_{i=1}^r v_i^n
  (p_{i,0} + n p_{i,1} + \cdots + n^k p_{i,k})$. By
  Theorems~\ref{thm:cfinite-zero} and~\ref{thm:poly-zero} we get that the
  $p_{i,j}$ have to be $0$. Therefore, also $v_i^n p_{i,j} = 0$ for all $i,j$.
  Then, for each $j=0,\dots,k$, we have ${v_1^np_{1,j} + \cdots + v_r^np_{1,j} = 0
  = q_j}$.
\end{proof}
\fi 

As $p$ is an algebraic relation, we have that $p'$ should be $0$
for all $n\in\N$. Proposition~\ref{prop:coeff-zero} then implies that the $q_i$
have to be $0$ for all $n\in\N$.
\begin{lemma}
  \label{lemma:cfinite}
  Let $q$ be of the form~\eqref{eq:qform}. Then ${q=0}$ for all ${n\in\N}$ if
  and only if ${q=0}$ for ${n\in\{0,\dots,\ell-1\}}$. \qed
\end{lemma}

\ifarxiv 
\begin{proof}
  The proof follows from Theorem~\ref{thm:finite-init-values} and from the fact
  that $q$ satisfies a C-finite recurrence of order $l$. To be more precise, the
  $u_{i,j}$ and $w_{i,j}^n$ satisfy a first-order C-finite recurrence: as
  $u_{i,j}$ is constant it satisfies a recurrence of the form $x(n+1)=x(n)$, and
  $w_{i,j}^n$ satisfies $x(n+1) = w_i x(n)$. Then, by
  Theorem~\ref{thm:cfinite-closure} we get that $w_{i,j}^n u_{i,j}$ is C-finite
  of order at most $1$, and $q$ is C-finite of order at most $\ell$.
\end{proof}
\fi 

Even though the $q_i$ contain exponential terms in $n$, it follows from
Lemma~\ref{lemma:cfinite} that the solutions for the $q_i$ being $0$ for all
$n\in\N$ can be described as a finite set of polynomial equality
constraints: 
Let $Q_i^j$ denote the polynomial constraint $w_{i,1}^ju_{i,1} + \cdots +
w_{i,\ell}^j u_{i,\ell} = 0$ for $q_i$ of the form~\eqref{eq:qform}, and let ${\CC_i
= \{Q_i^0,\dots,Q_i^{\ell-1}\}}$ be the associated clause set. Then the clause set
ensuring that $p$ is indeed an algebraic relation is given by
\begin{equation*}
  \Calg = \CC_0 \cup \cdots \cup \CC_k.
\end{equation*}
%


\begin{remark}
  Observe that Theorem~\ref{thm:finite-init-values} can be applied
  to~\eqref{eq:invpoly} directly, as $p'$ satisfies a C-finite
  recurrence. Then by the closure properties of C-finite recurrences, 
  the upper bound on the order of the recurrence which $p'$ satisfies 
  is given by $r =
  \sum_{i=0}^k 2^i \ell$. That is, by Theorem~\ref{thm:finite-init-values}, we
  would need to consider $p'$ with $n=0,\dots,r-1$, which yields a non-linear
  system with a degree of at least $r-1$. Note that $r$ depends on
  $2^i$,  which stems from the fact that $\seq{n}$ satisfies
  a recurrence of order $2$, and $n^i$ satisfies therefore a recurrence of order
  at most $2^i$. Thankfully, Proposition~\ref{prop:coeff-zero} allows us to only consider
  the coefficients of the $n^i$ and therefore lower the size of our constraints.\qed
\end{remark}

Having defined the clause sets $\Croots$, $\Ccoeff$, $\Cinit$ and
$\Calg$, we define our PCP as the union of these four clause
sets. Note that
the matrix $B$, the vector $A$, the polynomial $p$ and the multiplicities of the
symbolic roots $\vec{m} = m_1,\dots,m_t$ uniquely define the clauses discussed
above. We hence define our PCP to be the clause set
$\CC_{AB}^p(\vec{m})$ as follows:
\begin{equation}
  \CC_{AB}^p(\vec{m}) = \Croots \cup \Cinit \cup \Ccoeff \cup \Calg
\end{equation}
%

Recall that $\vec{a}$ and $\vec{b}$ are the symbolic entries in the matrices $A$
and $B$ in~\eqref{eq:nonparam:recsys}, $\vec{c}$ are the symbolic entries in the
$C_{ij}$ in~\eqref{eq:generalform}, and $\vec{\geom}$ are the symbolic eigenvalues
of $B$. We then have
$\Croots\polysin\K[\vec{\geom},\vec{b}]$,
${\Ccoeff\polysin\K[\vec{\geom},\vec{b},\vec{c}]}$,
${\Cinit\polysin\K[\vec{a},\vec{b},\vec{c}]}$ and
${\Ccoeff\polysin\K[\vec{\geom},\vec{c}]}$.
Hence ${\CC_{AB}^p(\vec{m})\polysin\K[\vec{\geom},\vec{a},\vec{b},\vec{c}]}$.

It is not difficult to see that the constraints in $\Calg$ determine
the
size of our PCP. 
As such, the degree and the number of terms
in the invariant have a direct impact on the size and the maximum degree of the
polynomials in our PCP. Which might not be obvious is that the
number of distinct symbolic roots influences the size and the maximum
degree of our PCP.
The more distinct roots are considered the higher is the number of terms in
\eqref{eq:qform}, and therefore more instances of \eqref{eq:qform}
have to be added to our PCP.

Let ${p\in\K[x_1,\dots,x_s,y_1,\dots,y_s]}$, ${B\in\K^{s\times s}}$ and
${A\in\K^{s\times 1}}$, and let ${m_1,\dots,m_t}$ be an integer partition of
$\deg_\geom(\chi_B(\geom))$. We then get the following theorem:

\begin{theorem}
  \label{thm:nonparam}
  The mapping ${\sigma:\{\vec{\geom},\vec{a},\vec{b},\vec{c}\}\rightarrow\K}$ is
  a solution of ${\CC_{AB}^p(\vec{m})}$ if and only if
  $p(\vec{x},x_1(0),\dots,x_s(0))$ is an algebraic relation for ${X_{n+1} =
  \sigma(B) X_n}$ with ${X_0 = \sigma(A)}$, and the eigenvalues of $\sigma(B)$
  are given by $\sigma(\geom_1),\dots,\sigma(\geom_t)$ with multiplicities
  $m_1,\dots,m_t$.\qed
\end{theorem}


\SetKwData{nothing}{nothing}\SetKwData{sat}{sat}
\SetKwFunction{intpart}{IntPartitions}\SetKwFunction{solve}{Solve}

From Theorem~\ref{thm:nonparam}, we then get Algorithm~\ref{alg:nonparam} for
synthesizing the C-finite recurrence representation of a non-parameterized
loop~\eqref{eq:loop}:  the function $\intpart(s)$ returns the set of all integer
partitions of an integer $s$; and $\solve(\CC)$ returns whether the clause set
$\CC$ is satisfiable and a model $\sigma$ if so. We note that the growth of the
number of integer partitions is subexponential, and so is the complexity
Algorithm~\ref{alg:nonparam}. A more precise complexity analysis of
Algorithm~\ref{alg:nonparam} is an interesting future work. 

\begin{algorithm}[tb]
\DontPrintSemicolon
\SetKwInOut{Input}{Input}\SetKwInOut{Output}{Output}
\Input{A polynomial $p\in\K[x_1,\dots,x_s,y_1,\dots,y_s]$.}
\Output{A vector $A\in\K^{s\times 1}$ and a matrix $B\in\K^{s\times s}$ s.t.~$p$
is an algebraic relation of $X_{n+1} = BX_{n}$ and $X_0 = A$, if such $A$ and
$B$ exist.}
\BlankLine
$A \gets (a_{i}) \in \K^{s\times1}$ \tcp{symbolic vector}
$B \gets (b_{ij}) \in \K^{s\times s}$ \tcp{symbolic matrix}
\For{$m_1,\dots,m_t\in \intpart(s)$}{
  $\sat, \sigma \gets \solve(\CC^p_{AB}(m_1,\dots,m_t))$\;
  \lIf{$\sat$}{
    \Return{$\sigma(A), \sigma(B)$}
  }
}

\caption{Synthesis of a non-parameterized C-finite recurrence system}
\label{alg:nonparam}
\end{algorithm}

Finally, based on  Theorem~\ref{thm:nonparam} and on the property that
the number of integer
partitions of a given integer is finite, we obtain the following
result: 

\begin{theorem}
  \label{thm:sound-complete}
  Algorithm~\ref{alg:nonparam} is sound, and complete w.r.t.~recurrence systems
  of size $s$.\qed
\end{theorem}

The completeness in Theorem~\ref{thm:sound-complete} is relative to systems of
size $s$ which is a consequence of the fact that we synthesize first-order
recurrence systems. That is, there exists a recurrence system of order~$>1$ and
size $s$ with an algebraic relation $p\in\K[x_1,\dots,x_s]$, but there exists no
first-order system of size $s$ where $p$ is an algebraic relation. 

The precise characterization of non-parameterized loops by non-parameterized
C-finite recurrence systems implies soundness and completeness for
non-parameterized loops from Theorem~\ref{thm:sound-complete}.

\begin{example}\label{ex:nonparam}
  We showcase our procedure in Algorithm~\ref{alg:nonparam} by synthesizing a loop for the invariant ${x=2y}$.
  That is, the polynomial constraint is given by ${p = x-2y \in \K[x,y]}$ and we
  want to find a recurrence system of the following form:
  \begin{equation}
    \label{ex:eq:general-rec}
    \begin{pmatrix}
      x(n+1) \\
      y(n+1)
    \end{pmatrix}=
    \begin{pmatrix}
      b_{11} & b_{12} \\
      b_{21} & b_{22}
    \end{pmatrix}
    \begin{pmatrix}
      x(n) \\
      y(n)
    \end{pmatrix}
    \qquad\qquad
    \begin{pmatrix}
      x(0) \\
      y(0)
    \end{pmatrix}=
    \begin{pmatrix}
      a_1 \\
      a_2
    \end{pmatrix}
  \end{equation}
  The characteristic polynomial of $B$ is then given by
  %
    $\chi_B(\geom) = \geom^2 - b_{11}\geom - b_{22}\geom - b_{12}b_{21}
    + b_{11}b_{22}$ 
  %
  where its roots define the closed form system. Since we cannot determine the
  actual roots of $\chi_B(\geom)$ we have to fix a set of symbolic roots. The
  characteristic polynomial has two -- not necessarily distinct -- roots: Either
  $\chi_B(\geom)$ has two distinct roots $\geom_1,\geom_2$ with multiplicities
  ${m_1=m_2=1}$, or a single root $\geom_1$ with multiplicity ${m_1=2}$. Let us
  consider the latter case. The first clause set we define is $\Croots$ for
  ensuring that $B$ is invertible (i.e.~$\geom_1$ is nonzero), and that
  $\geom_1$ is indeed a root of the characteristic polynomial with multiplicity
  $2$. That is, $\chi_B(\geom) = (\geom - \geom_1)^2$ has to hold for all
  $\geom\in\K$, and bringing everything to one side yields 
  \begin{equation*}
    (b_{11} + b_{22} - 2\geom_1) \geom + b_{12}b_{21} - b_{11}b_{22} + \geom_1^2 = 0.
  \end{equation*}
  %
  We then get the following clause set: 
  \begin{equation*}
    \Croots = \{ b_{11} + b_{22} - 2\geom_1 = 0, b_{12}b_{21} - b_{11}b_{22} + \geom_1^2 = 0, \geom_1\neq0 \}
  \end{equation*}
  As we fixed the symbolic roots, the general closed form system is
  of the form 
  \begin{equation}
    \label{ex:eq:general-cf}
    \begin{pmatrix}
      x(n) \\ y(n)
    \end{pmatrix} =
    \begin{pmatrix}
      c_1 \\ c_2
    \end{pmatrix} \geom_1^n +
    \begin{pmatrix}
      d_1 \\ d_2
    \end{pmatrix} \geom_1^n n
  \end{equation}
  By substituting into the recurrence system we get:
  \begin{equation*}
    \begin{pmatrix}
      c_1 \\ c_2
    \end{pmatrix} \geom_1^{n+1} +
    \begin{pmatrix}
      d_1 \\ d_2
    \end{pmatrix} \geom_1^{n+1} (n+1) =
    \begin{pmatrix}
      b_{11} & b_{12} \\
      b_{21} & b_{22}
    \end{pmatrix}
    \left(\begin{pmatrix}
      c_1 \\ c_2
    \end{pmatrix} \geom_1^n {+}
    \begin{pmatrix}
      d_1 \\ d_2
    \end{pmatrix} \geom_1^n n \right)
  \end{equation*}
  By further simplifications and re-ordering of terms  we then obtain:
  %
  %
  \begin{align*}
    0 = \begin{pmatrix}
      c_1\geom_1+d_1\geom_1 - b_{11}c_1-b_{12}c_2 \\ c_2\geom_1+d_2\geom_1 - b_{21}c_1-b_{22}c_2
    \end{pmatrix} \geom_1^{n}
    +
    \begin{pmatrix}
      d_1\geom_1 - b_{11}d_1-b_{12}d_2 \\ d_2\geom_1 - b_{21}d_1-b_{22}d_2
    \end{pmatrix} \geom_1^{n} n
  \end{align*}
  Since this equation has to hold for $n\in\N$ we get the following clause set:
  \begin{align*}
    \Ccoeff = \{
      &c_1\geom_1+d_1\geom_1 - b_{11}c_1-b_{12}c_2=0, 
      c_2\geom_1+d_2\geom_1 - b_{21}c_1-b_{22}c_2=0, \\
      &d_1\geom_1 - b_{11}d_1-b_{12}d_2=0, 
      d_2\geom_1 - b_{21}d_1-b_{22}d_2=0
    \}
  \end{align*}
  For defining the relationship between the closed forms and
  the initial values, we set \eqref{ex:eq:general-cf} with $n=i$ to be
  equal to the $i^\kth$ unrolling of \eqref{ex:eq:general-rec} for $i=0,1$:
  \begin{equation*}
    \begin{pmatrix}
      c_1 \\ c_2
    \end{pmatrix} = 
    \begin{pmatrix}
      a_1 \\ a_2
    \end{pmatrix}
    \qquad
    \begin{pmatrix}
      c_1 \\ c_2
    \end{pmatrix} \geom_1 +
    \begin{pmatrix}
      d_1 \\ d_2
    \end{pmatrix} \geom_1 = 
    \begin{pmatrix}
      b_{11} & b_{12} \\
      b_{21} & b_{22}
    \end{pmatrix}
    \begin{pmatrix}
      a_1 \\ a_2
    \end{pmatrix}
  \end{equation*}
  The resulting constraints for defining the initial values are then given by
  \[
    \Cinit = \{ c_1-a_1=0, c_1\geom_1 + d_1\geom_1 - b_{11}a_1 - b_{12}a_2 = 0,
                c_2-a_2=0, c_2\geom_1 + d_2\geom_1 - b_{21}a_1 - b_{22}a_2 = 0 \}.
  \]
  Eventually, we want to restrict the solutions such that $x-2y=0$ is an algebraic
  relation for our recurrence system. That is, by substituting the closed forms
  into $x(n)-2y(n)=0$ we get
  \[
    0 = x(n) - 2y(n) 
       = c_1\geom_1^n + d_1\geom_1^nn - 2(c_2\geom_1^n + d_2\geom_1^nn)
       = \underbrace{\left(c_1-2c_2\right)\geom_1^n}_{q_0} + \underbrace{\left(\left( d_1-2d_2 \right)\geom_1^n\right)}_{q_1} n
  \]
  where $q_0$ and $q_1$ have to be $0$ since the above equation has to hold for
  all $n\in\N$. Then, by applying Lemma~\ref{lemma:cfinite} to $q_0$ and $q_1$,
  we get the following clauses:
  \begin{equation*}
    \Calg = \{ c_1-2c_2=0, d_1-2d_2=0 \}
  \end{equation*}
 Our PCP is then the union of $\Croots$, $\Ccoeff$, $\Cinit$ and
  $\Calg$. Two possible solutions for our PCP,  and therefore of
  the synthesis problem, are given by the following loops:
  \begin{equation*}
    \begin{tabular}{l}
    $(x, y) \gets (2, 1)$\\
    $\WHILE~true~\DO$
    $(x, y) \gets (x+2, y+1)$
      $\END$\\
    \end{tabular}
    \qquad\quad
    \begin{tabular}{l}
      $(x, y) \gets (2, 1)$\\
      $\WHILE~true~\DO$
      $(x, y) \gets (2x, 2y)$
      $\END$\\
    \end{tabular}
  \end{equation*}
  Note that both loops above have mutually independent updates. Yet, the second
  one induces geometric sequences and requires handling exponentials of $2^n$.\qed
\end{example}

\ifarxiv 
\subsection{Synthesizing Parameterized Loops}
\label{sec:synth:param}
\input{mfcs_param}
\fi 

\section{Implementation and Experiments}\label{sec:experiments}

Our approach to algebra-based loop synthesis is implemented in the tool
\absynth{} which is available at {\absynthurl}. Inputs to \absynth{} are
conjunctions of polynomial equality constraints, representing a loop invariant.
As a result, \absynth{} derives a program that is partially correct with respect
to the given invariant.

Loop synthesis in \absynth{} is reduced to solving PCPs. These PCPs are
expressed in the quantifier-free fragment of non-linear real arithmetic
(\texttt{QF\_NRA}). We used \absynth{} in conjunction with the SMT solvers
\tool{Yices}~\cite{Yices} and \tool{Z3}~\cite{Z3} for solving the PCPs and
therefore synthesizing loops. For instance, the loops depicted in
Figures~\ref{fig:Dafny:b} and~\ref{fig:Dafny:c}, and in
Example~\ref{ex:nonparam} are synthesized automatically using \absynth{}.

\subparagraph*{Optimizing and Exploring the Search Space.}
\label{sec:synth:opt}
\absynth{} implements additional constraints to restrict the search space of
solutions to loop synthesis. Namely, \absynth{} (i) avoids trivial
loops/solutions and (ii) restricts the shape of $B$ to be triangular or
unitriangular. The latter allows \absynth{}  to synthesize loops whose loop
variables are not mutually dependent on each other. We note that such a pattern
is a very common  programming paradigm -- all benchmarks from
Table~\ref{tab:experiments} in Appendix~\ref{sec:appendix:experiments} satisfy
such a pattern. Yet, as a consequence of restricting the shape of $B$, the order
of the variables in the recurrence system matters. That is, we have to consider
all possible variable permutations for ensuring completeness
w.r.t.~(uni)triangular matrices.

\absynth{} however supports an iterative approach for exploring the solution
space. One can start with a small recurrence system and a
triangular/unitriangular matrix $B$, and then stepwise increase the
size/generality of the system. Our initial results from
Table~\ref{tab:experiments} in Appendix~\ref{sec:appendix:experiments}
demonstrate the practical use of our approach to loop synthesis: all examples
could be solved in reasonable time. 


\input{related}


\section{Conclusions}
\label{sec:conclusion}

We proposed a syntax-guided synthesis procedure for synthesizing loops from a
given polynomial loop invariant.  
We consider loop templates and
use reasoning over recurrence equations modeling the loop behavior. The key
ingredient of our work comes with translating the loop synthesis problem into a
polynomial constraint problem and showing that this constraint problem precisely
captures all solutions to  the loop synthesis problem. 
We implemented our work and evaluated on a number of
academic examples. 
Understanding and encoding the best optimization measures for loop
synthesis is an interesting line for future work. 



\bibliography{references}

\newpage
\appendix
\section{Appendix}
\label{appendix}

\subsection{Examples and Experiments}
\label{sec:appendix:experiments}

\newcommand{\TO}{-}
\newcommand{\UN}{}

\newcommand{\full}{\textsc{fu}}
\newcommand{\upper}{\textsc{up}}
\newcommand{\uni}{\textsc{un}}

\begin{table}
  \centering
  \def\arraystretch{.95}
  \begin{tabular}{@{}lrrrrrrrrrrrrrrrr@{}}
    \toprule
    \multirow{2}{*}{Instance} & \multirow{2}{*}{{\textsc{s}}} & \multirow{2}{*}{\textsc{i}} & \multirow{2}{*}{\textsc{d}} & \multirow{2}{*}{\textsc{c}} & \phantom{x} & \multicolumn{3}{c}{\tool{Yices}} & \phantom{x} & \multicolumn{3}{c}{\tool{Z3}} & \phantom{x} & \multicolumn{3}{c}{\tool{Z3*}} \\ 
    \cmidrule{7-9} \cmidrule{11-13} \cmidrule{15-17}
    &&&&&& \uni & \upper & \full && \uni & \upper & \full && \uni & \upper & \full \\
    \midrule
    \texttt{add1}*      & 5 & 1 &  5 & 173 && $932$ \UN & $921$ \UN  & \TO        && $117$ \UN & \TO       & \TO       && $22$ \UN & $726$ \UN & \TO       \\
    \texttt{add2}*      & 5 & 1 &  5 & 173 && $959$ \UN & $861$ \UN  & \TO        && $115$ \UN & \TO       & \TO       && $22$ \UN & $109$ \UN & \TO       \\
    \texttt{cubes}      & 5 & 3 &  6 &  94 && \TO       & \TO        & \TO        && $116$ \UN & $114$ \UN & \TO       && $18$ \UN & $496$ \UN & $575$ \UN \\
    \texttt{double1}    & 3 & 1 &  4 &  29 && $114$ \UN & $112$ \UN  & $3882$ \UN && $113$ \UN & $111$ \UN & $113$ \UN && $13$ \UN &  $21$ \UN &  $63$ \UN \\
    \texttt{double2}    & 3 & 1 &  3 &  24 && $110$ \UN & $106$ \UN  & $1665$ \UN && $115$ \UN & $106$ \UN & $115$ \UN && $13$ \UN &  $18$ \UN &  $40$ \UN \\
    \texttt{eucliddiv}* & 5 & 1 &  5 & 185 && $213$ \UN & $537$ \UN  & \TO        && $114$ \UN & $115$ \UN & \TO       && $19$ \UN &  $73$ \UN & \TO       \\
    \texttt{intcbrt}*   & 5 & 2 & 12 & 262 && \TO       & \TO        & \TO        && $117$ \UN & $116$ \UN & \TO       && $22$ \UN &  $83$ \UN & $469$ \UN \\
    \texttt{intsqrt1}   & 4 & 2 &  6 &  53 && \TO       & \TO        & \TO        && $113$ \UN & $108$ \UN & $114$ \UN && $15$ \UN &  $19$ \UN & \TO       \\
    \texttt{intsqrt2}*  & 4 & 1 &  6 & 104 && $105$ \UN & $1164$ \UN & \TO        && $113$ \UN & $111$ \UN & $115$ \UN && $15$ \UN &  $27$ \UN &  $37$ \UN \\
    \texttt{petter1}    & 3 & 1 &  4 &  29 && $112$ \UN & $116$ \UN  & \TO        && $114$ \UN & $113$ \UN & $113$ \UN && $15$ \UN &  $18$ \UN &  $32$ \UN \\
    \texttt{square}     & 3 & 1 &  4 &  29 && $112$ \UN & $112$ \UN  & \TO        && $112$ \UN & $114$ \UN & $117$ \UN && $13$ \UN &  $17$ \UN &  $26$ \UN \\
    \texttt{dblsquare}  & 3 & 1 &  4 &  30 && $109$ \UN & $105$ \UN  & \TO        && $105$ \UN & $105$ \UN & $110$ \UN && $12$ \UN &  $17$ \UN &  $26$ \UN \\
    \texttt{sum1}       & 4 & 2 &  6 &  53 && $617$ \UN & \TO        & \TO        && $108$ \UN & $112$ \UN & $113$ \UN && $17$ \UN &  $24$ \UN &  $99$ \UN \\
    \texttt{sum2}       & 5 & 3 &  6 &  82 && \TO       & \TO        & \TO        && $220$ \UN & $112$ \UN & \TO       && $20$ \UN & $516$ \UN & \TO       \\
    \bottomrule
  \end{tabular}
  \vskip1ex
  %
  %
  \def\arraystretch{1}
  \begin{tabular}{llcll}
    \textsc{s} & size of the recurrence system & \phantom{xx} &
      {}* & parameterized system \\
    \textsc{i} & number of polynomial invariants & &
      - & timeout (60 seconds) \\
    \textsc{d} & maximum monomial degree of constraints\\
    \textsc{c} & number of constraints\\
  \end{tabular}
  \vspace{.5em}
  \caption{Benchmark results in milliseconds}
  \label{tab:experiments}
\end{table}


Table~\ref{tab:experiments} summarizes our experimental results. 
The experiments were performed on a machine with a 2.9 GHz Intel Core i5 and 16
GB LPDDR3 RAM, and for each instance a timeout of 60 seconds was set. The
results are given in milliseconds, and only include the time needed for solving
the constraint problem as the time needed for constructing the constraints is
neglectable. We used the SMT solvers \tool{Yices}~\cite{Yices} (version 2.6.1)
and \tool{Z3}~\cite{Z3} (version 4.8.6) to conduct our experiments. In
Table~\ref{tab:experiments}, the columns \tool{Yices} and \tool{Z3} correspond
to the results where the respective solver is called as an external program with
and SMTLIB 2.0 file as input; column \tool{Z3*} shows the results
where our improved, 
direct interface (C\texttt{++} API) was used to call \tool{Z3}.

Our  benchmark set consists of invariants for loops from the invariant
generation literature. Note that the benchmarks \texttt{cubes} and
\texttt{double2} in Table~\ref{tab:experiments} are those from
Figure~\ref{fig:Dafny} and Example~\ref{ex:nonparam}, respectively. A further
presentation of a selected set of our benchmarks is given in
Appendix~\ref{sec:appendix:examples}.

Our work supports an iterative approach for exploring the solution space of
loops to be synthesized. One can start with a small recurrence system and a
triangular/unitriangular matrix $B$, and then stepwise increase the
size/generality of the system. The columns \uni{} and \upper{} in
Table~\ref{tab:experiments} show the results where the coefficient matrix $B$ is
restricted to be  upper unitriangular and upper triangular respectively.
\full{} indicates that no restriction on $B$ was set.

Note that the running time of Algorithm~\ref{alg:nonparam} heavily depends on
the order of which the integer partitions and the variable permutations are
traversed. Therefore, in order to get comparable results, we fixed the integer
partition and the variable permutation. That is, for each instance, we enforced
that $B$ has just a single eigenvalue, and we fixed a variable ordering where we
know that there exists a solution with a unitriangular matrix $B$. Hence, there
exists at least one solution which all cases -- \uni{}, \upper{} and \full{} --
have in common. Furthermore, for each instance we added constraints for avoiding
trivial solutions, i.e.~loops inducing constant sequences.

\subsection{Examples of Synthesized Loops}
\label{sec:appendix:examples}


We took loops from the invariant generation literature and computed their
invariants. Our benchmark set consists of these generated invariants. For each
example in Figures~\ref{fig:experiments:a}-\ref{fig:experiments:e}, we first
list the original loop and then give the first loop synthesized by our
work in
combination with \tool{Yices} and \tool{Z3} respectively.

Observe that in most cases our work was able to derive the original loop --
apart from the initial values -- with either \tool{Z3} or \tool{Yices}.

\newcommand{\firstcol}{0.27\textwidth}
\newcommand{\secondcol}{0.30\textwidth}
\newcommand{\thirdcol}{0.41\textwidth}

\begin{figure*}
\begin{center}
\begin{minipage}[t]{\firstcol}
\begin{lstlisting}[escapechar=\%]
  # \textbf{eucliddiv}
  # Original loop
  r, q, y = x0, 0, y0
  while true
    r = r - y
    q = q + 1
  end
  %~%
\end{lstlisting}
\end{minipage}
\begin{minipage}[t]{\secondcol}
\begin{lstlisting}[escapechar=\%]
  # \textbf{eucliddiv}
  # Solver: Yices
  r, q, y = x0, 0, y0
  while true
    r = r - q - y
    q = q + 1
    y = y - 1
  end
\end{lstlisting}
\end{minipage}
\begin{minipage}[t]{\thirdcol}
\begin{lstlisting}[escapechar=\%]
  # \textbf{eucliddiv}
  # Solver: Z3
  r, q, y = x0 - 1/2 y0, 1/2, y0
  while true
    r = r - q - 1/2 y + 1/2
    q = q + 1/2
    y = y - 1
  end
\end{lstlisting}
\end{minipage}
\end{center}

\vspace*{-1em}
\caption{Example \texttt{eucliddiv} with input \texttt{x0 == y0*q+r}}
\label{fig:experiments:a}
\end{figure*}

\begin{figure*}
\begin{center}
\begin{minipage}[t]{\firstcol}
\begin{lstlisting}
  # \textbf{square}
  # Original loop
  a, b = 0, 0
  while true
    a = a + 2b + 1
    b = b + 1
  end
\end{lstlisting}
\end{minipage}
\begin{minipage}[t]{\secondcol}
\begin{lstlisting}
  # \textbf{square}
  # Solver: Yices
  a, b = 0, 0
  while true
    a = a - 2b + 1
    b = b - 1
  end
\end{lstlisting}
\end{minipage}
\begin{minipage}[t]{\thirdcol}
\begin{lstlisting}
  # \textbf{square}
  # Solver: Z3
  a, b = 1/16, -1/4
  while true
    a = a + 2b + 1
    b = b + 1
  end
\end{lstlisting}
\end{minipage}

\end{center}

\vspace*{-1em}
\caption{Example \texttt{square} with input \texttt{a == b\^{}2}}
\label{fig:experiments:b}
\end{figure*}

\begin{figure*}
\begin{center}
\begin{minipage}[t]{\firstcol}
\begin{lstlisting}
  # \textbf{sum1}
  # Original loop
  a, b, c = 0, 0, 1
  while true
    a = a + 1
    b = b + c
    c = c + 2
  end
\end{lstlisting}
\end{minipage}
\begin{minipage}[t]{\secondcol}
\begin{lstlisting}
  # \textbf{sum1}
  # Solver: Yices
  a, b, c = 1/2, 1/4, 2
  while true
    a = a - 1/2
    b = b - 1/2 c + 3/4
    c = c - 1
  end
\end{lstlisting}
\end{minipage}
\begin{minipage}[t]{\thirdcol}
\begin{lstlisting}
  # \textbf{sum1}
  # Solver: Z3
  a, b, c = -5/8, 25/64, -1/4
  while true
    a = a + 1
    b = b + c
    c = c + 2
  end
\end{lstlisting}
\end{minipage}

\end{center}

\vspace*{-1em}
\caption{Example \texttt{sum1} with input \texttt{1+2a == c \&\& 4b == (c-1)\^{}2}}
\label{fig:experiments:c}
\end{figure*}

\begin{figure*}
\begin{center}
\begin{minipage}[t]{\firstcol}
\begin{lstlisting}
  # \textbf{intsqrt2}
  # Original loop
  y, r = 1/2 a0, 0
  while true
      y = y - r
      r = r + 1
  end
\end{lstlisting}
\end{minipage}
\begin{minipage}[t]{\secondcol}
\begin{lstlisting}
  # \textbf{intsqrt2}
  # Solver: Yices
  y, r = 1/2 a0, 0
  while true
    y = y + r - 1
    r = r - 1
  end
\end{lstlisting}
\end{minipage}
\begin{minipage}[t]{\thirdcol}
\begin{lstlisting}
  # \textbf{intsqrt2}
  # Solver: Z3
  y, r  = 1/2 a0 - 5/32, -1/4
  while true
      y = y - r
      r = r + 1
  end
\end{lstlisting}
\end{minipage}

\end{center}

\vspace*{-1em}
\caption{Example \texttt{intsqrt2} with input \texttt{a0+r == r\^{}2+2y}}
\label{fig:experiments:d}
\end{figure*}

\begin{figure*}
\begin{center}
\begin{minipage}[t]{0.29\textwidth}
\begin{lstlisting}
  # \textbf{intcbrt}
  # Original loop
  x, r, s = a0, 1, 13/4
  while true
    x = x - s
    s = s + 6r + 3
    r = r + 1
  end
\end{lstlisting}
\end{minipage}
\begin{minipage}[t]{0.71\textwidth}
\begin{lstlisting}
  # \textbf{intcbrt}
  # Solver: Z3
  x, s, r = 34/64 + a0, 7/16, -1/4
  while true
    x = x - s
    s = s + 6r + 3
    r = r + 1
  end
\end{lstlisting}
\end{minipage}

\end{center}

\vspace*{-1em}
\caption{Example \texttt{intcbrt} with input \texttt{1/4+3r\^{}2 == s \&\& 1+4a0+6r\^{}2 == 3r+4r\^{}3+4x}}
\label{fig:experiments:e}
\end{figure*}

\end{document}

%% file: mfcs_param.tex
We now extend the loop synthesis approach from Section~\ref{sec:synth:nonparam}
to an algorithmic approach synthesizing parameterized loops, that is, loops
which satisfy a loop invariant for arbitrary input values. Let us first consider
the following example motivating the synthesis problem of parameterized loops. 

\begin{example}
  \label{ex:eucliddiv}
  We are interested to synthesize a loop implementing Euclidean division over
  $x,y\in\K$. Following the problem specification of~\cite{knuth}\footnote{for
  $x,y\in\K$ we want to compute $q,r\in\K$ such that $x = yq + r$ holds}, a
  synthesized loop performing Euclidean division satisfies the polynomial
  invariant ${p = \ini{x} - \ini{y}q - r = 0}$, where $\ini{x}$ and $\ini{y}$
  denote the initial values of $x$ and $y$ before the loop. It is clear, that
  the synthesized loop should be parameterized with respect to $\ini{x}$ and
  $\ini{y}$.
  With this setting, input to our synthesis approach is the invariant ${p =
  \ini{x} - \ini{y}q - r = 0}$. A recurrence system performing Euclidean
  division and therefore satisfying the algebraic relation $\ini{x} - \ini{y}q -
  r$ is then given by $X_{n+1} = BX_n$ and $X_0 = A$ with a corresponding closed
  form system $X_n = A + Cn$ where: 
  \begin{equation*}
    X_n = \begin{pmatrix}x(n) \\ r(n) \\ q(n) \\ y(n) \\ t(n)\end{pmatrix}
    \quad
    A = \begin{pmatrix}\ini{x} \\ \ini{x} \\ 0 \\ \ini{y} \\ 1\end{pmatrix}
    \quad
    B =
    \begin{pmatrix}
      1 & 0 & 0 & 0 & 0 \\
      0 & 1 & 0 & -1 & 0 \\ 
      0 & 0 & 1 & 0 & 1 \\
      0 & 0 & 0 & 1 & 0 \\
      0 & 0 & 0 & 0 & 1
    \end{pmatrix}
    \quad
    C = \begin{pmatrix} 0 \\ -\ini{y} \\ 1 \\ 0 \\ 0 \end{pmatrix}
  \end{equation*}
  Here, the auxiliary variable $t$ plays the role of the constant $1$, and $x$
  and $y$ induce constant sequences. When compared to non-parameterized C-finite
  systems/loops, note that the coefficients in the above closed forms, as well
  as the initial values of variables, are functions in the parameters $\ini{x}$
  and $\ini{y}$.\qed
\end{example}

Example~\ref{ex:eucliddiv} illustrates that the parameterization has the effect
that we have to consider parameterized closed forms and initial values. For
non-parameterized loops we have that the coefficients in the closed forms are
constants, whereas for parameterized systems we have that the coefficients are
functions in the parameters -- the symbolic initial values of the sequences. In
fact, we have linear functions since the coefficients are obtained by solving a
linear system (see Example~\ref{ex:fibonacci}).

As already mentioned, the parameters are a subset of the symbolic initial values
of the sequences. Therefore, let ${I = \{k_1,\dots,k_r\}}$ be a subset of the
indices $\{1,\dots,s\}$. We then define $\ini{X} =
\begin{pmatrix} \ini{x}_{k_1} & \cdots & \ini{x}_{k_r} &
1\end{pmatrix}^\intercal$ where $\ini{x}_{k_1},\dots,\ini{x}_{k_r}$ denote the
parameters. Then, instead of~\eqref{eq:nonparam:recsys}, we get
\begin{equation}
  \label{eq:param:recsys}
  X_{n+1} = BX_n \qquad X_0 = A\ini{X}
\end{equation}
as the implicit representation of our recurrence system where the entries of
${A\in\K^{s\times r+1}}$ are defined as
\begin{equation*}
  a_{ij} = 
  \begin{cases}
    1 & i = k_j \\
    a_{ij}~\text{symbolic} & i \notin I \\
    0 & \text{otherwise}
  \end{cases}
\end{equation*}
and, as before, we have $B\in\K^{s\times s}$. Intuitively, the complex looking
construction of $A$ makes sure that we have $x_i(0)=\ini{x}_i$ for $i\in I$.
\begin{example}
  For the vector ${X_0 = \begin{pmatrix} x_1(0) & x_2(0) & x_3(0)
  \end{pmatrix}^\intercal}$, the set ${I=\{1,3\}}$ and therefore ${\ini{X} =
  \begin{pmatrix} \ini{x}_1 & \ini{x}_3 & 1 \end{pmatrix}^\intercal}$, we get
  the following matrix:
  \begin{equation*}
    A = \begin{pmatrix}
      1 & 0 & 0 \\
      a_{21} & a_{22} & a_{23} \\
      0 & 1 & 0
    \end{pmatrix}
  \end{equation*}
  Thus, $x_1(0)$ and $x_3(0)$ are set to $\ini{x}_1$ and $\ini{x}_3$
  respectively, and $x_2(0)$ is a linear function in $\ini{x}_1$ and
  $\ini{x}_3$.\qed
\end{example}
In addition to the change in the representation of the initial values, we also
have a change in the closed forms. That is, instead of~\eqref{eq:generalform} we
get
\begin{equation*}
  X_n = \sum_{i=1}^t \sum_{j=1}^{m_i} C_{ij}\ini{X} \geom_i^n n^{j-1}
\end{equation*}
as the general form for the closed form system with $C_{ij}\in\K^{s\times r+1}$.
Then $\Croots$, $\Cinit$, $\Ccoeff$ and $\Calg$ are defined analogously to
Section~\ref{sec:synth:nonparam}, and similar to the non-parameterized case we
define $\CC^p_{AB}(\vec{m},\vec{\ini{x}})$ as the union of those clause sets.
The polynomials in $\CC^p_{AB}(\vec{m},\vec{\ini{x}})$ are then in
$\K[\vec{\geom},\vec{a},\vec{b},\vec{c},\vec{\ini{x}}]$. Then, for each
$\vec{\geom},\vec{a},\vec{b},\vec{c} \in\K$ satisfying the clause set for all
$\vec{\ini{x}}\in\K$ gives rise to the desired parameterized loop, that is, we
have to solve an $\exists\forall$ problem. However, since all constraints
containing $\vec{\ini{x}}$ are polynomial equality constraints, we apply
Theorem~\ref{thm:inifite-zeros}:
Let ${p\in\K[\vec{\geom},\vec{a},\vec{b},\vec{c},\vec{\ini{x}}]}$ be a
polynomial such that ${p= p_1 q_1 + \dots + p_k q_k}$ with
${p_i\in\K[\vec{\ini{x}}]}$ and $q_i$ monomials in
$\K[\vec{\geom},\vec{a},\vec{b},\vec{c}]$. Then, Theorem~\ref{thm:inifite-zeros}
implies that the $q_i$ have to be $0$.


We therefore define the following operator $\decompose_{\vec{x}}(p)$ for
collecting the coefficients of all monomials in $\vec{x}$ in the polynomial $p$:
Let $p$ be of the form $q_0 + q_1x + \cdots + q_kx^k$, $P$ a clause and let
$\CC$ be a clause set, then:
\begin{align*}
  \decompose_{\vec{y},x}(p) &= 
  \begin{cases}
    \{q_0=0,\dots,q_k=0\} &\text{if $\vec{y}$ is empty} \\
    \decompose_{\vec{y}}(q_0)\cup\cdots\cup\decompose_{\vec{y}}(q_k) &\text{otherwise}
  \end{cases}\\
  \decompose_{\vec{y}}(P) &= 
  \begin{cases}
    \decompose_{\vec{y}}(p) &\text{if $P$ is a unit clause $p=0$} \\
    \{P\} &\text{otherwise}
  \end{cases}\\
  \decompose_{\vec{y}}(\CC) &= \bigcup_{P\in\CC} \decompose_{\vec{y}}(P)
\end{align*}
We then have ${\decompose_{\vec{\ini{x}}}(\CC^p_{AB}(\vec{m},\vec{\ini{x}}))
\polysin \K[\vec{\geom},\vec{a},\vec{b},\vec{c}]}$. Moreover, for
${p\in\K[x_1,\dots,x_s,y_1,\dots,y_s]}$, matrices $A$,$B$ and $\ini{X}$ as
in~\eqref{eq:param:recsys}, and an integer partition ${m_1,\dots,m_t}$ of
$\deg_\geom(\chi_B(\geom))$ we get the following theorem:

\begin{theorem}
  \label{thm:param}
  The map ${\sigma:\{\vec{\geom},\vec{a},\vec{b},\vec{c}\}\rightarrow\K}$ is
  a solution of
  ${\decompose_{\vec{\ini{x}}}(\CC^p_{AB}(\vec{m},\vec{\ini{x}}))}$
  if and only if $p(\vec{x},x_1(0),\dots,x_s(0))$ is an algebraic relation for
  $X_{n+1} = \sigma(B) X_n$ with $X_0 = \sigma(A)\ini{X}$, and
  $\sigma(\geom_1),\dots,\sigma(\geom_t)$ are the eigenvalues of $\sigma(B)$
  with multiplicities $m_1,\dots,m_t$.\qed
\end{theorem}

Theorem~\ref{thm:param} gives rise to an algorithm analogous to
Algorithm~\ref{alg:nonparam}. Furthermore, we get an analogous soundness and
completeness result as in Theorem~\ref{thm:sound-complete} which implies
soundness and completeness for parameterized loops.

\begin{example}
  We illustrate the construction of the constraint problem for
  Example~\ref{ex:eucliddiv}. For reasons of brevity, we consider a simplified
  system where the variables $r$ and $x$ are merged. The new invariant is then
  $\ini{r} = \ini{y}q + r$ and the parameters are given by $\ini{r}$ and
  $\ini{y}$. That is, we consider a recurrence system of size $4$ with sequences
  $y$, $q$ and $r$, and $t$ for the constant $1$.  As a consequence we have that
  the characteristic polynomial $B$ is of degree $4$, and we fix the symbolic
  root $\geom_1$ with multiplicity $4$. For simplicity, we only show how to
  construct the clause set $\Calg$.
  
  With the symbolic roots fixed we get the following template for the closed
  form system: Let ${X_n = \begin{pmatrix}r(n) & q(n) & y(n) &
  t(n)\end{pmatrix}^\intercal}$ and ${V = \begin{pmatrix} \ini{r} & \ini{y} & 1
  \end{pmatrix}^\intercal}$, and let ${C,D,E,F\in\K^{4\times 3}}$ be symbolic
  matrices. Then the closed form is given by
  \begin{align*}
    X_n &= \left(CV + DVn + EVn^2 + FVn^3\right) \geom_1^n
  \end{align*}
  and for the initial values we get
  \begin{align*}
    X_0 &=
    \begin{pmatrix}
      1 & 0 & 0 \\ a_{21} & a_{22} & a_{23} \\ 0 & 1 & 0 \\ a_{41} & a_{42} & a_{43}
    \end{pmatrix}V.
  \end{align*}
  By substituting the closed forms into the invariant ${r(0) - y(0) q(n) - r(n)
  = 0}$ and rearranging we get:
  \begin{align*}
    0 = \ini{r} &- \left(c_{21}\ini{r}\ini{y} - c_{22}\ini{y}^2 - c_{23}\ini{y} - c_{11}\ini{r} - c_{12}\ini{y} - c_{13}\right) \geom_1^n \\
                &- \left(d_{21}\ini{r}\ini{y} + d_{22}\ini{y}^2 + d_{23}\ini{y} - d_{11}\ini{r} - d_{12}\ini{y} - d_{13}\right) \geom_1^n n \\
                &- \left(e_{21}\ini{r}\ini{y} + e_{22}\ini{y}^2 + e_{23}\ini{y} - e_{11}\ini{r} - e_{12}\ini{y} - e_{13}\right) \geom_1^n n^2 \\
                &- \left(f_{21}\ini{r}\ini{y} + f_{22}\ini{y}^2 + f_{23}\ini{y} - f_{11}\ini{r} - f_{12}\ini{y} - f_{13}\right) \geom_1^n n^3
  \end{align*}
  Since the above equation should hold for all $n\in\N$ we get: 
  \begin{align*}
    \left(\ini{r}\right) 1^n - \left(c_{21}\ini{r}\ini{y} - c_{22}\ini{y}^2 - c_{23}\ini{y} - c_{11}\ini{r} - c_{12}\ini{y} - c_{13}\right) \geom_1^n &= 0 \\
    \left(d_{21}\ini{r}\ini{y} + d_{22}\ini{y}^2 + d_{23}\ini{y} - d_{11}\ini{r} - d_{12}\ini{y} - d_{13}\right) \geom_1^n &= 0 \\
    \left(e_{21}\ini{r}\ini{y} + e_{22}\ini{y}^2 + e_{23}\ini{y} - e_{11}\ini{r} - e_{12}\ini{y} - e_{13}\right) \geom_1^n &= 0\\
    \left(f_{21}\ini{r}\ini{y} + f_{22}\ini{y}^2 + f_{23}\ini{y} - f_{11}\ini{r} - f_{12}\ini{y} - f_{13}\right) \geom_1^n &= 0
  \end{align*}
  Then, by applying Lemma~\ref{lemma:cfinite}, we get:
  \begin{align*}
    \ini{r} - \left(c_{21}\ini{r}\ini{y} - c_{22}\ini{y}^2 - c_{23}\ini{y} - c_{11}\ini{r} - c_{12}\ini{y} - c_{13}\right) &= 0 \\
    \ini{r} - \left(c_{21}\ini{r}\ini{y} - c_{22}\ini{y}^2 - c_{23}\ini{y} - c_{11}\ini{r} - c_{12}\ini{y} - c_{13}\right) \geom_1 &= 0 \\
    d_{21}\ini{r}\ini{y} + d_{22}\ini{y}^2 + d_{23}\ini{y} - d_{11}\ini{r} - d_{12}\ini{y} - d_{13} &= 0\\
    e_{21}\ini{r}\ini{y} + e_{22}\ini{y}^2 + e_{23}\ini{y} - e_{11}\ini{r} - e_{12}\ini{y} - e_{13} &= 0\\
    f_{21}\ini{r}\ini{y} + f_{22}\ini{y}^2 + f_{23}\ini{y} - f_{11}\ini{r} - f_{12}\ini{y} - f_{13} &= 0
  \end{align*}
  Finally, by applying the operator $\decompose_{\ini{y},\ini{r}}$, we get the
  following constraints for $\Calg$:
  \begin{alignat*}{5}
    c_{21} &={ }& 1-c_{11} &={ }& c_{22} &={ }& c_{23} + c_{12} &={ }& c_{13} &= 0 \\
    \geom_1c_{21} &={ }& 1-\geom_1c_{11} &={ }& \geom_1c_{22} &={ }& \geom_1\left(c_{23} + c_{12}\right) &={ }& \geom_1c_{13} &= 0 \\
    d_{21} &={ }& d_{11}   &={ }& d_{22} &={ }& d_{23} + d_{12} &={ }& d_{13} &= 0 \\
    e_{21} &={ }& e_{11}   &={ }& e_{22} &={ }& e_{23} + e_{12} &={ }& e_{13} &= 0 \\
    f_{21} &={ }& f_{11}   &={ }& f_{22} &={ }& f_{23} + f_{12} &={ }& f_{13} &= 0
  \end{alignat*}\qed
\end{example}

%% file: related.tex
\section{Related Work}\label{sec:related}
%

\subparagraph*{Synthesis.} To the
best of our knowledge, existing synthesis approaches are restricted to linear
invariants, see e.g.~\cite{GulwaniPOPL10}, whereas our work supports
loop synthesis from non-linear polynomial properties. 
In the setting of counterexample-guided synthesis -- CEGIS~\cite{Alur18,Solar09, DilligPLDI18,SolarICML19},
input-output examples satisfying a specification $S$ are used to synthesize a candidate program $P$
that is consistent with the given inputs. Correctness
of the candidate program $P$ with respect to $S$ is then checked using
verification approaches, in particular using SMT-based reasoning. If
verification fails, a counterexample is generated as an input to $P$ that
violates $S$. This counterexample is then used in conjunction with the previous
set of input-outputs to revise synthesis and generate a new candidate program
$P$.
Unlike these methods, input
specifications to our approach are relational (invariant) properties describing
all, potentially infinite input-output examples of interest. Hence, we do not
rely on interactive refinement of our input but work with a precise
characterization of the set of input-output values of the program to be
synthesized. Similarly to sketches~\cite{Solar09,SolarICML19}, we consider loop templates restricting the
search for solutions to synthesis. Yet, our templates support non-linear
arithmetic (and hence multiplication), which is not yet the case
in~\cite{SolarICML19,DilligPLDI18}.
We precisely characterize the set of all programs
satisfying our input specification, and as such, our approach does not exploit
learning to refine program candidates. On the other hand, our programming model
is more restricted than~\cite{SolarICML19,DilligPLDI18} in various aspects:
we only handle simple loops and only consider numeric data types and operations.

The programming by example approach of~\cite{GulwaniPOPL11} learns programs from
input-output examples and relies on lightweight interaction to refine the
specification of programs to be specified. The approach has further been
extended in~\cite{GulwaniICLR18} with machine learning, allowing to learn
programs from just one (or even none) input-output example by using a simple
supervised learning setup. Program synthesis from input-output examples is shown
to be successful for recursive programs~\cite{GulwaniCAV13}, yet synthesizing
loops and handling non-linear arithmetic is not yet supported by this
line of research.
Our work does not learn programs from
observed input-output examples, but uses loop invariants to fully characterize
the intended behavior of the program to be synthesized. Our technique precisely
characterizes the solution space of loops to be synthesized by a system of
algebraic recurrences, and hence we do not rely on statistical models supporting
machine learning. 

A related approach to our work is tackled
in~\cite{KuncakEMSOFT13}, where a fixed-point implementation for an approximated
real-valued polynomial specification is presented, by combining 
genetic programming~\cite{GP08} 
with abstract
interpretation~\cite{CousotC77} to estimate and refine the (floating-point) error
bound of the inferred fixed-point implementation. While the underlying abstract
interpreter is precise for linear expressions, precision of the synthesis is lost
in the presence of non-linear arithmetic. Unlike~\cite{KuncakEMSOFT13},
we consider polynomial specification in the abstract algebra of real-closed
fields and do not address challenges rising from machine
reals.

\subparagraph*{Algebraic Reasoning.} When compared to works on generating polynomial
invariants~\cite{Rodriguez-CarbonellK07,HumenbergerJK17,KincaidCBR18,Worrell18},
the only common aspect between these works and our synthesis
method is the use of linear recurrences to capture the  functional
behavior of program loops. Yet, our work is conceptually different
than~\cite{Rodriguez-CarbonellK07,HumenbergerJK17,KincaidCBR18,Worrell18},
as 
we reverse engineer invariant generation and do not rely on the ideal structure/Zariski closure of
polynomial invariants. We do not use ideal theory nor Gr\"obner bases
computation to generate invariants from loops; rather, we generate
loops from invariants by formulating and solving PCPs. 